\documentclass{natures}
\usepackage{amsmath}
\usepackage{mathtools}
\usepackage{amssymb}
\usepackage{caption}
\usepackage{graphicx}

\usepackage{tabularx}		\newcolumntype{Y}{>{\centering\arraybackslash}X}
\usepackage{amsfonts}
\usepackage[ansinew]{inputenc}
\usepackage[usenames,dvipsnames]{pstricks}
\usepackage{subfigure}
\usepackage{multirow,bigdelim}
\usepackage{array}

\DeclareCaptionLabelSeparator{period-newline}{ $\vert $ }
\title{Dark trions and biexcitons in ${\rm WS_2}$ and ${\rm WSe_2}$ made bright by e-e scattering}

\author{Mark Danovich$^{1}$, Viktor Z\'{o}lyomi$^1$ \& Vladimir I. Fal'ko$^1$}

\begin{document}

\maketitle

\begin{affiliations}
	\item National Graphene Institute, University of Manchester, Booth St E, Manchester M13 9PL, UK
\end{affiliations}

\begin{abstract}
The direct band gap character and large spin-orbit splitting of the valence band edges (at the K and K' valleys) in monolayer transition metal dichalcogenides have put these two-dimensional materials under the spot-light of intense experimental and theoretical studies\cite{tmdcs_opto,cao,coup_spin_valley,opt-gen,valley-pol,wang,kdotp}. 
In particular, for Tungsten dichalcogenides it has been found\cite{heinz,exchange_split,kdotp,wang} that the sign of spin splitting of conduction band edges makes ground state excitons radiatively inactive (dark) due to spin and momentum mismatch between the constituent electron and hole. One might similarly assume that the ground states of charged excitons and biexcitons in these monolayers are also dark. Here, we show that the intervalley $(K\leftrightarrows K')$ electron-electron scattering mixes bright and dark states of these complexes, and estimate the radiative lifetimes in the ground states of these ``semi-dark'' trions and biexcitons to be $\sim 10~{\rm ps}$, and analyse how these complexes appear in the temperature-dependent photoluminescence spectra of ${\rm WS_2}$ and ${\rm WSe_2}$ monolayers.
\end{abstract}

The truly 2D nature of TMDCs enhances the effects of Coulomb interaction\cite{2dscreen,chernikov}, resulting in charge complexes such as excitons\cite{timothy,mak,longname_excitons,op_spec}, trions\cite{timothy} and biexcitons\cite{biex} with binding energies that are orders of magnitude larger compared to conventional semiconductors such as GaAs. These complexes, which dominate the optical response of these materials, are comprised of spin/valley polarised electrons and holes residing at the corners K and K' of the hexagonal Brillouin zone (BZ), where the selection rules of optical transitions require the same spin and valley states of the involved electrons at the conduction and valence band edges. As a result, the opposite spin projections of the conduction ($c$) and valence ($v$) band edges, found in monolayers of ${\rm WS_2}$ and ${\rm WSe_2}$, makes ground state excitons in these 2D crystals dark\cite{heinz}, so that their radiative transition would require help from defects, phonons\cite{danovich_auger} or magnetic field\cite{heinz_b,potem_b}.

\begin{figure}
	\centering
	\includegraphics[width=0.85\textwidth]{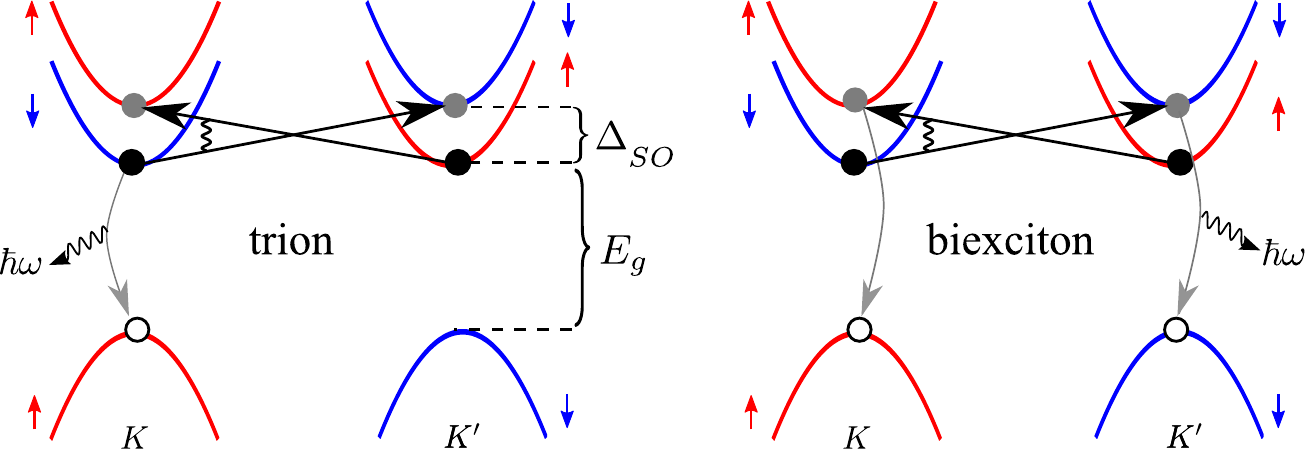}
	\captionsetup{labelfont=bf}
	\captionsetup{labelsep=period-newline}
	\caption{ 
		{\bf Intervalley electron-electron scattering process.}\\
		{\small Schematics of the band structures of ${\rm WX_2}$ near the $K,K'$ points of the BZ, and the intervalley scattering process that mixes dark and bright states of trions (T) and biexcitons (B). $E_g$ is the band gap and $\Delta_{SO}$ stands for the conduction band spin splitting. Due to the large spin-orbit splitting in the valence band, the valence band is shown only for the higher-energy spin-polarised states.}}
		\vspace{0.5cm}
	\label{fig:dark}
\end{figure}

Applying the spin and valley selection rules to ground state trions and biexcitons might imply that these charge complexes are dark, too. In the `dark' ($d$) state both electrons are in the bottom spin-orbit split states of $c$-band, whereas in the state to be `bright' ($b$), one of the electrons has to be in the excited spin-split state. Here, we show that an intervalley scattering\cite{vexcitons,dery} of the $c$-band electrons mixes dark and bright states of complexes (Fig.~\ref{fig:dark}), hence transferring some optical strength from $b$- to $d$-states and making dark state `semi-dark'. For the resulting recombination line of such semi-dark complexes, we find that it is shifted downwards in energy (relative to the bright exciton line) by $\sim2\Delta_{SO}$, twice the $c$-band spin-orbit splitting.

With the reference to Fig.~\ref{fig:dark}, the basis of trion, T (biexciton, B) states, $T_{\sigma_c\tau_c,\sigma_{c'}\tau_{c'}}^{\sigma_v\tau_v}$ ( $B_{\sigma_c\tau_c,\sigma_{c'}\tau_{c'}}^{\sigma_v\tau_v,\sigma_{v'}\tau_{v'}}$), can be described by spin, $\sigma=\uparrow,\downarrow$ and valley, $\tau=K,K'$ quantum numbers of their constituent $c$- ad $v$-band states. In these notations, dark ground state exciton complexes $T_d$ ($B_d$) are $T_{\downarrow K,\uparrow K'}^{\uparrow K}$ and $T_{\downarrow K,\uparrow K'}^{\downarrow K'}$ ($B_{\downarrow K,\uparrow K'}^{\uparrow K,\downarrow K'}$), and the excited states $T_{\uparrow K,\downarrow K'}^{\uparrow K}$ and $T_{\uparrow K,\downarrow K'}^{\downarrow K'}$ ($B_{\uparrow K,\downarrow K'}^{\uparrow K,\downarrow K'}$) are bright, $T_b$ ($B_b$) (Supplementary material S1). 
These states are mixed by the intervalley interaction illustrated by a sketch in Fig.~\ref{fig:dark}, 
\begin{equation}
H_{iv}=\frac{\hbar^2\chi}{2m_c}\sum_{\sigma,\tau}\int d^2\vec{r} \Psi^{\dagger}_{c,\sigma,-\tau}(\vec{r})\Psi^{\dagger}_{c,-\sigma,\tau}(\vec{r})\Psi_{c,-\sigma,-\tau}(r)\Psi_{c,\sigma,\tau}(\vec{r}).
\label{hi}
\end{equation}
Here, $\Psi_{c,\sigma,\tau}(\vec{r})$ are the conduction band electron field operators. The large momentum transfer between two electrons changing their valley states is determined by their Coulomb interaction at the unit cell scale, parametrised by a dimensionless factor $\chi$. We estimate the size of this factor using both a tight-binding model and density functional theory (DFT). 
For the tight-binding model, we use the DFT calculated orbital decomposition to construct the Bloch states at the Brillouin zone corners, and we use a 3D Coulomb potential for the interaction between electrons. As the $c$-band states at the $K/K'$ points are primarily composed\cite{wang,kdotp} of the metal $5d_{z^2}$ orbitals centred at the lattice sites $\vec{R}$ of metallic atoms in TMDC lattice, $\phi(\vec{r}-\vec{R})$, which we use to construct the tight-binding model Bloch states, to find  
\begin{equation}
\chi=\frac{m_c}{m}\frac{A}{a_B} |C|^4 \sum_{\vec{R}}e^{i\vec{K}\cdot\vec{R}}\int
d^3 \vec{r}_1d^3\vec{r}_2 \frac{|\phi(\vec{r}_1)|^2|\phi(\vec{r}_2)|^2}{|\vec{r}_2-\vec{r}_1+\vec{R}|}.
\label{eq:mat_el}
\end{equation}
Here, $\vec{K}=(\frac{4\pi}{3a_0},0)$ with $a_0$ the lattice constant of ${\rm WX_2}$,
$A=\frac{\sqrt{3}}{2} a_0^2$ is the unit cell area, $m_c$ is the $c$-band electron effective mass, $m$ is the free electron mass, $a_B$ is the Bohr radius, and $C$ is the transition metal $5d_{z^2}$ orbital amplitude in the $c$-band edge at the $K$ point (Supplementary material S2.2).  Similarly, we evalutaed $\chi$ from wave functions obtained using DFT implemented in the local density approximation and VASP\cite{vasp} code (neglecting spin-orbit coupling). We 
used a plane-wave basis corresponding to $600~{\rm eV}$ cutoff energy and a $12 \times 12$ grid of k-points in the 2D Brillouin zone. We also had to employ periodic boundary conditions in the $z$-direction; for this reason we used a large inter-layer distance of $20~{\rm \AA}$ to mimic the limit of an isolated monolayer. The form factor was calculated by post-processing the DFT wave functions, by taking the matrix element of the bare Coulomb interaction between the initial and final states of the scattering process (Supplementary material S2.1). These two calculations have returned values of the intervalley scattering factor $\chi$, as listed in Table~\ref{table:rates}. 
\begin{table}
	\captionsetup{labelfont=bf}
	\captionsetup{labelsep=period-newline}
	\caption{{\bf Material parameters.} 
		\\{\small Listed are the effective $c-$ and $v$-band electron masses, $c$-band spin-orbit splitting, 2D screening length, bright exciton energy, trion binding energy, biexciton binding energy, and the velocity related to the off diagonal momentum matrix element.}}
	\label{table:data}
	\centering
	\begin{tabular}{lccccccccc}
		\hline\hline
		& $\frac{m_c}{m}$~[\citenum{kdotp}] &  $\frac{m_v}{m}$~[\citenum{kdotp}]& $\Delta_{SO}$~[\citenum{kdotp}]& $A$~[\citenum{kdotp}] & $r_*$~[\citenum{timothy}]  & $E_{X_b}$~[\citenum{excitonse}] & $\epsilon_T$~[\citenum{binding}] & $\epsilon_{B}$~[\citenum{binding}] & $\frac{v}{c}$~[\citenum{kdotp}]  \\ 
		& & & $[{\rm meV]}$ & $[{\rm nm^2}]$ & $[{\rm nm}]$ & $[{\rm eV}]$ & $[{\rm meV}]$ & $[{\rm meV}]$   \\
		\hline
		${\rm WS_2}$ & $0.26$ & $-0.35$  & $32$ & $8.65$ & $3.8$  & $2$ & $34$ & $24$ & $1.7\times 10^{-3}$ 
		\\
		${\rm WSe_2}$ & $0.28$ & $-0.36$ & $37$ & $9.38$ & $4.5$ & $1.7$ & $31$ & $20$ & $1.6\times 10^{-3}$ 
		\\
		\hline\hline
	\end{tabular}
\vspace{0.8cm}
\end{table}
In the basis of $[|d\rangle;|b\rangle]$ of dark and bright states of trions, $[T_{\downarrow K,\uparrow K'}^{\uparrow K};T_{\uparrow K,\downarrow K'}^{\uparrow K}]$ and
$[T_{\downarrow K,\uparrow K'}^{\downarrow K'};T_{\uparrow K,\downarrow K'}^{\downarrow K'}]$, 
or biexcitons $[B_{\downarrow K,\uparrow K'}^{\uparrow K,\downarrow K'};B_{\uparrow K,\downarrow K'}^{\uparrow K,\downarrow K'}]$, the coupling in Eq.~(\ref{hi}) leads to the mixing described by a $2\times2$ matrix
\begin{align}
&H=\left(\begin{matrix}
E_b^{T/B} & \mu_{T/B} \\
\mu^*_{T/B} & E_d^{T/B}
\end{matrix}\right), 
\quad  \mu_T=\frac{\hbar^2\chi}{m_c}g_T,  \quad \mu_B=\frac{\hbar^2\chi}{m_c}g_B,
\\
&E_b^T=2E_g+2\Delta_{SO}-\epsilon_X-\epsilon_T+\delta',\nonumber
\\
&E_d^T=2E_g-\epsilon_X-\epsilon_T+\delta,\nonumber
\\
&E_b^B=2E_g+2\Delta_{SO}-\epsilon_X-\epsilon_B+2\delta',\nonumber
\\
&E_d^B=2E_g-\epsilon_X-\epsilon_B+2\delta.\nonumber
\end{align}
Where $E_g$ is the band gap, $\epsilon_X, \epsilon_T$, and $\epsilon_B$ are the exciton, trion, and biexciton binding energies, respectively, and $\delta, \delta'$ stand for the intravalley and intervalley electron-hole exchange\cite{trionfine}, $\delta\approx 6~{\rm meV}$, which we will neglect in the following calculations. Note that the effective masses of the $c$-band spin split bands differ by\cite{kdotp} $\sim 30\--40\%$ with the lower bands having the higher effective electron mass. This results in slightly higher binding energies for the dark ground state charge complexes compared to the excited states, resulting in a larger value for their energy difference $E_b-E_d$.
The mixing parameter $\mu\equiv \langle b|H_{iv}|d\rangle=\frac{\hbar^2\chi}{m_c}\int \prod\limits_{i} d^2\vec{r}_i|\Phi_{T/B}|^2\delta(\vec{r}_e-\vec{r}_{e'})$, (where $\Phi_{T/B}$ stands for the wave function of the trion or biexciton and $i=e,e',h,(h')$, is determined by the electron-electron contact pair densities\cite{ganchev} in the trion, $g_T$ and biexciton, $g_B$.

\begin{figure}
	\centering
	\includegraphics[width=0.9\textwidth]{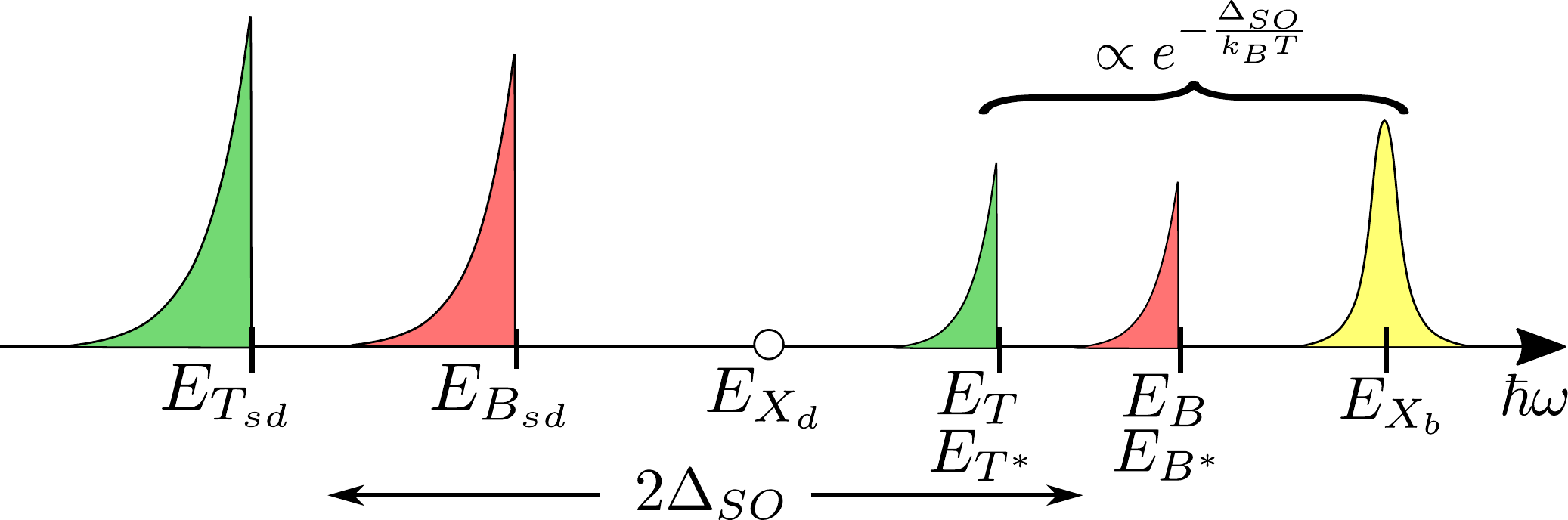}
	\captionsetup{labelfont=bf}
	\captionsetup{labelsep=period-newline}
	\caption{ {\bf Low temperature photoluminescence spectrum of ${\rm WX_2}$.}\\
		Sketch of the low temperature ($k_BT<\Delta_{SO}$) photoluminescence spectrum of ${\rm WX_2}$ including the bright exciton, dark and bright trions (green) and dark and bright biexcitons (red). The excited bright trions and excitons are denoted by $T^*$ and $B^*$.
		The dark exciton ($X_d$) energy is marked as a reference point $E_{X_d}=E_{X_b}-\Delta_{SO}$.}
	\vspace{0.5cm}
	\label{fig:spectrum}
\end{figure}

The mixing of the dark and bright states results in a slight shift of their energies and, most importantly, in a finite radiative decay rate, $\tau_{sd}^{-1}$ of the semi-dark (\textit{sd}) trions (\textit{T}) and biexcitons (\textit{B}), 
\begin{align}
&\frac{1}{\tau_{sd}}\approx\left(1-\frac{1}{\sqrt{1+\left(\frac{\mu_{T/B}}{\Delta_{SO}}\right)^2}}\right)\frac{\alpha_{T/B}}{2}\tau^{-1}_X, 
\\
&\nonumber \frac{1}{\tau_X}=\frac{8\pi}{\hbar}\frac{e^2}{\hbar c}\frac{\hbar^2v^2}{E_{X_b}}|\Phi_X(0)|^2,
\end{align}
where $\tau_X^{-1}$ is the radiative decay rate of the bright exciton\cite{excitons,wang_rad,basko-radiative}, determined by the electron-hole overlap factor $|\Phi_X(0)|^2$ ($\Phi_X(r_{eh})$ is the envelope wave function describing relative motion of the electron and hole in the exciton), $v$ is the velocity related to the off diagonal momentum matrix element. The values of the  factors $\alpha_T=\frac{1}{2}$ and $\alpha_B=\frac{2}{3}$ have been estimated based on the following consideration. As the exciton's binding energy is significantly larger than that of the trion or biexciton, these bound complexes can be viewed as strongly-bound, with an additional weakly bound electron in the case of a trion, or an exciton in the case of a biexciton. For a trion, this results in a reduction of the recombining electron-hole contact pair density by a factor of 2, as the 
hole is shared between the two electrons such that the recombining electron (which has the right spin projection), will be near it only half of the time. In the case of the biexciton, the recombining electron will be part of the time near the other hole and part of the time the other electron will be near the hole, giving a factor of $1/3$, however in this case both holes can recombine radiatively with a proper electron producing an additional factor of 2, hence, giving $\alpha_B=\frac{2}{3}$. 
The resulting values for the lifetimes (using the material parameters in Table~\ref{table:data}) are summarized in Table~\ref{table:rates}.
\begin{table}
	\captionsetup{labelfont=bf}	
	\captionsetup{labelsep=period-newline}
	\caption{{\bf Radiative lifetimes and scattering matrix elements.}\\
		{\small Listed are the Intervalley scattering parameter $\chi$ calculated using DFT and tight binding (TB) model and the corresponding trion and biexciton mixing parameters $\mu_{T/B}$ obtained using the electron-electron contact pair densities calculated in ref.~\citenum{ganchev} using quantum Monte Carlo, shown as DFT [TB], and the radiative lifetimes of the bright exciton, semi-dark trion and biexciton.}}
	\label{table:rates}
	\centering
	\begin{tabular}{lccccccc}
		\hline \hline
		& $\chi_{DFT}$ & $\chi_{TB}$ & $\mu_T$ & $\mu_B$  &$\tau_X$ & $\tau_{sd}(T)$ & $\tau_{sd}(B)$ \\
		& & & $[{\rm meV}]$ & $[{\rm meV}]$ & $[{\rm ps}]$ & $[{\rm ps}]$& $[{\rm ps}]$ \\ 
		\hline
		${\rm WS_2}$ & $1.0$ & $1.6$ &  $18~[29]$ & $13~[21]$&$0.25$ & $7.7~[3.9]$ & $10~[4.5]$\\ 
		${\rm WSe_2}$ & $1.3$ & $2.0$ & $19~[30]$ & $14~[22]$ &$0.26$ & $9.1~[4.7]$ & $12~[5.7]$ \\ 
		\hline\hline
	\end{tabular}
\vspace{0.8cm}
\end{table}
The mixing of the dark and bright states produces photoluminescence lines shown schematically in Fig.~\ref{fig:spectrum}. The emitted photon energies of these lines are determined by both the binding energies and the shake-up into the higher-energy spin-split $c$-band in the final state,
\begin{align}
&E_{X_b}=E_g+\Delta_{SO}-\epsilon_X,
\\
&E_{T_{sd}/B_{sd}}\approx E_{X_b}-\epsilon_{T/B}-2\Delta_{SO}, \nonumber
\\
&E_{T/B}\approx E_{X_b}-\epsilon_{T/B}. \nonumber
\end{align}
 Being the ground states, the semi-dark trion and biexcitons $(T_{sd}, B_{sd})$ do not require an activation and therefore should appear in the spectrum even at low temperatures. In contrast, the bright states do require thermal activation, resulting in a $e^{-\Delta E/k_BT}$ temperature dependence of their lines intensities. For the bright exciton, trion $[T^{\uparrow K}_{\uparrow K, \uparrow K'}; T^{\downarrow K'}_{\downarrow K, \downarrow K'}]$ and biexciton $[B^{\uparrow K,\downarrow K'}_{\uparrow K, \uparrow K'}; B^{\uparrow K,\downarrow K'}_{\downarrow K, \downarrow K'}]$ we have $\Delta E\approx\Delta_{SO}$, while for the excited mixed dark and bright trion ($T^*$) $[T^{\uparrow K}_{\uparrow K, \downarrow K'}; T^{\downarrow K'}_{\uparrow K, \downarrow K'}]$ and biexciton  ($B^*$) $B^{\uparrow K, \downarrow K'}_{\uparrow K, \downarrow K'}$, $\Delta E\approx 2\Delta_{SO}$. Also, the presence of a final state electron or exciton results in an antisymmetric line shape with a cutoff due to the recoil kinetic energy of the remaining electron or exciton that shifts the emission line to a lower energy. A typical recoil kinetic energy is $\frac{m_X}{m_c}k_BT$ for the trions and $k_B T$ for biexcitons, with $k_B$ the Boltzmann constant, $m_X$ the exciton mass, and $m_c$ the $c$-band electron effective mass.

In conclusion, we have shown that intervalley electron-electron scattering makes ``dark'' ground state trions and biexcitons in Tungsten dichalcogenides ${\rm WS_2}$ and ${\rm WSe_2}$ optically active, with a lifetime $\tau_{T/B}\sim 10~{\rm ps}$, to compare with a sub-ps lifetime of bright excitons in 2D TMDCs.

\begin{addendum}
	\item The authors would like to thank I. Aleiner, T. Heinz, M. Potemski, M. Syniszewski, A. Tartakovski and X. Xu for useful discussions.
\end{addendum}

\newcommand{\rep}{\emph{rep} }
\newcommand{\irep}{\emph{irrep} }	
\newcommand{\beginsupplement}{%
	\setcounter{table}{0}
	\renewcommand{\thetable}{S\arabic{table}}%
	\setcounter{figure}{0}
	\renewcommand{\thefigure}{S\arabic{figure}}%
	\renewcommand{\thesection}{S\arabic{section}}%
	\renewcommand{\theequation}{S\arabic{equation}}%
}
\clearpage
\begin{center}
{\bf \large Supplementary material}	
\end{center}
	\beginsupplement{
		\section{Group theory analysis of excitons, trions and biexcitons in Tungsten dichalcogenides}
		\label{secgroup}
		
		\subsection{Introduction}
		
		Group theory allows to utilize the symmetry properties of the Hamiltonian in order to gain insight into selection rules for microscopic processes in quantum systems. As a starting point, the eigenstates of the Hamiltonian are classified according to the irreducible representations (IrReps) of the symmetry group, in our case the point group $C_{3h}$. In monolayer TMDCs, DFT calculations\cite{wangs,kdotps} (see also S2.1) have found that band edges of monolayer ${\rm WS_2}$ and ${\rm WSe_2}$ are found at the two inequivalent corners, $K$ and $K'$ of the Brillouin zone. Hence, for the sake of their classification we consider the extended point group\cite{baskovs, danovich_augers}, $C_{3v}^{''}=C_{3v}+tC_{3v}+t^2C_{3v}$, where $t$ are translations by a lattice vector.
		This enables us to treat states of excitons and complexes at $K$, $K'$ and zero momentum in the same fashion.
		The character table and product table for the IrReps of the extended point group $C_{3v}''$ are given in Tables~S1, S2, respectively.
		DFT calculations\cite{wangs,kdotps}  (see also S2.1) have also found that at the $K$ and $K'$ valleys, the orbital composition of the Bloch states is dominated by the $z\rightarrow -z$ symmetric $d$-orbitals ($d_0$ for the $c$-band and  $d_{\pm 2}$ for the $v$-band in the two valleys) of transition metal, allowing to classify the $c$ and $v$-band Bloch states at the $K$ and $K'$ valleys as transforming according to the two dimensional IrReps of the extended point group, $E_1'$ and $E_2'$, respectively.
		
		\begin{table}
			\centering
			\captionsetup{labelfont=bf}
			\caption{\small{{\bf $C_{3v}''$ character table.}\\
					Character table for the irreducible representations (IrRep) of the extended point group $C_{3v}''$, and their correspondence to the conduction ($c$) and valence ($v$) band electrons states.}}
			\label{c3v''}
			\begin{tabular}{lcccccccc}
				\hline\hline
				$C_{3v}''$ & $E$ & $t,t^2$ & $2C_3$ & $9\sigma_v$ & $2tC_3$ & $2t^2C_3$ & \\
				\hline
				$A_1$ & 1 & 1 & 1 & 1 & 1 & 1 & \\ 
				$A_2$ & 1 & 1 & 1 & -1 & 1 & 1  & \\ 
				$E$ & 2 & 2 & -1 & 0 & -1 & -1 & \\
				\hline
				$E_1'~(c)$ & 2 & -1 & -1 & 0 & 2 & -1 &\\ 
				$E_2'~(v)$ & 2 & -1 & 2 & 0 & -1 & -1 & \\ 
				$E_3'$ & 2 & -1 & -1 & 0 & -1 & 2 &\\
				\hline \hline
			\end{tabular} 
		\end{table}
		
		\begin{table}
			\centering
			\captionsetup{labelfont=bf}
			\caption{\small{{\bf $C_{3v}''$ product table.}\\
					{Product table for the irreducible representations of the extended point group $C_{3v}''$.}}}
			\label{prod}
			\begin{tabular}{lcccccc}
				\hline\hline
				$C_{3v}''$ & $A_1$ & $A_2$ & $E$ & $E_1'$ & $E_2'$ & $E_3'$\\
				\hline
				$A_1$ & $A_1$ & $A_2$ & $E$ & $E_1'$ & $E_2'$ & $E_3'$ \\ 
				$A_2$ &  $A_2$& $A_1$ & $E$ & $E_1'$ & $E_2'$ & $E_3'$ \\ 
				$E$ & $E$ & $E$ & $A_1\oplus A_2\oplus E$ & $E_2'\oplus E_3'$ & $E_1'\oplus E_3'$ & $E_1'\oplus E_2'$ \\ 
				\hline
				$E_1'~(c)$ &$E_1'$ & $E_1'$ & $E_2'\oplus E_3'$& $A_1\oplus A_2\oplus E_1'$ & $E\oplus E_3'$ & $E\oplus E_2'$ \\ 
				$E_2'~(v)$ &$E_2'$& $E_2'$ & $E_1'\oplus E_3'$ & $E\oplus E_3'$ & $A_1\oplus A_2\oplus E_2'$ & $E\oplus E_1'$ \\ 
				$E_3'$ & $E_3'$ & $E_3'$ & $E_1'\oplus E_2'$ & $E\oplus E_2'$ & $E\oplus E_1'$ & $A_1\oplus A_2\oplus E_3'$\\
				\hline\hline
			\end{tabular} 
		\vspace{1cm}
		\end{table}
		
		Using classification of the single electron states, we consider excitons, trions, and biexcitons. For this, we take direct products of the corresponding IrReps, and, then, apply the product rules for the IrReps of $C_{3v}''$, shown in Table~\ref{prod}. This group theory analysis enables us to identify excitonic basis states that can be mixed by the intervalley e-e scattering, leading to the class of semi-dark trions and biexcitons discussed in the main text.
				
		\subsection{Excitons}
		The exciton states transform according to the direct product representation of the $c$- and $v$-band states given by 
		\begin{equation}
		E_1'\otimes E_2'=E\oplus E_3'.
		\end{equation}
		The 2D IrRep $E$ corresponds to the intravalley excitons with both electron and hole residing in either the $K$ or $K'$ valleys, and the 2D IrRep $E_3'$ corresponds to the intervalley excitons with the electron and hole residing in opposite valleys making the exciton dark due to momentum mismatch. By further introducing the spin projections of the electron and hole, we have for each representation two possible total spin projections, $|S_z|=1$ corresponding to dark excitons due to spin conservation, and $S_z=0$ corresponding to bright exciton states.
		Using the notation introduced in the text for trions and biexcitons, the $E$ IrRep dark intravalley exciton states are given by $[X^{\uparrow K}_{\downarrow K}; X^{\downarrow K'}_{\uparrow K'}]$ with $|S_z|=1$, and the bright intravalley excitonic states by $[X^{\uparrow K}_{\uparrow K}; X^{\downarrow K'}_{\downarrow K'}]$ with $|S_z|=0$. Similarly, for the intervalley excitons transforming according to $E_3'$, which are dark due to momentum conservation, we have $[X^{\uparrow K}_{\uparrow K'}; X^{\downarrow K'}_{\downarrow K}]$ with $S_z=0$, and $[X^{\uparrow K}_{\downarrow K'}; X^{\downarrow K'}_{\uparrow K}]$ with $S_z=1$, being dark due to both spin and momentum conservation.
		
		\subsection{Trions}
		Next we classify the trion states composed of two electrons and a hole. The strongly bound trion states require the two-electron wave function to be symmetric with respect to exchanging the electrons coordinates and the two electrons to have different spin/valley indices corresponding to a singlet state, as obtained in ref.~\citenum{ganchevs} using Monte Carlo calculations. The two-electron state transforms according to the direct product of the $c$-band electrons representations given by 
		\begin{equation}
		E_1'\otimes E_1'=A_1\oplus A_2\oplus E_1'.
		\end{equation}
		According to Table~\ref{c3v''}, the symmetric combination of the two electrons transforms according to $A_1$ or $E_1'$. The identity representation corresponds to both electrons residing in opposite valleys, while the 2D IrRep $E_1'$ corresponds to both electrons residing in the same valley $K$ or $K'$. Next, to obtain the representation of the trion we include the hole state $E_2'$ and take the direct product of the two electrons and the hole. This gives in the first case 
		\begin{equation}
		A_1\otimes E_2'=E_2', 
		\end{equation}
		corresponding to the hole residing in either the $K$ or $K'$ valleys and the electrons residing in opposite valleys. Including the spin projection this corresponds to the following trion states, $[T^{\uparrow K}_{\downarrow K, \uparrow K'}; T^{\downarrow K}_{\downarrow K, \uparrow K'}]$ which are the semi-dark singlet ground state trions, and $[T^{\uparrow K}_{\uparrow K, \downarrow K'}; T^{\downarrow K'}_{\uparrow K, \downarrow K'}]$ which are the excited bright trion singlet states. As the excited bright and semi-dark trion states both transform according to the same $E_2'$ IrRep, the two states can be mixed through the electron-electron intervalley scattering introduced in the main text, which transforms as the identity representation. The bright trion triplet states with both electrons in opposite valleys also transform according to the $E_2'$ IrRep and are given by $[T^{\uparrow K}_{\uparrow K, \uparrow K'}; T^{\downarrow K'}_{\downarrow K, \downarrow K'}]$, and the dark trion triplet states (due to spin conservation) are given by $[T^{\uparrow K}_{\downarrow K, \downarrow K'}; T^{\downarrow K'}_{\uparrow K, \uparrow K'}]$. In the second case, choosing for the two-electron representation the $E_1'$ IrRep,
		\begin{equation}
		E_1'\otimes E_2'=E\oplus E_3'. 
		\end{equation}
		Here, $E$ corresponds to states with the two electrons and hole residing in the same valley $K$ or $K'$. Requiring the electrons to have opposite spin projections gives the following bright trion states $[T^{\uparrow K}_{\uparrow K, \downarrow K}; T^{\downarrow K'}_{\uparrow K', \downarrow K'}]$. $E_3'$ corresponds to the two electrons residing in the same valley while the hole is in the opposite valley, giving the dark trion states (due to momentum conservation) $[T^{\downarrow K'}_{\uparrow K, \downarrow K}; T^{\uparrow K}_{\uparrow K', \downarrow K}]$.
		
		\subsection{Biexcitons}
		The bound biexciton states are composed of a spatially symmetric wave function for the two electrons and for the two holes. This corresponds to the IrReps $A_1\oplus E_1'$ for the two electrons, and $A_1\oplus E_2'$ for the two holes. Taking the direct product of the two-electron and two-hole states gives the possible representations of the biexciton states
		\begin{equation}
		(A_1\oplus E_1')\otimes(A_1\oplus E_2')=A_1\oplus E_1' \oplus E\oplus E_3' \oplus E_2'. 
		\end{equation}
		The states transforming according to the IrRep $E$ correspond to both electrons and both holes residing in the same valley, similarly the $E_3'$ IrRep corresponds to both electrons residing in the same valley and both holes residing in the opposite valley to the electrons, and finally $E_2'$ corresponds to both electrons residing in opposite valleys, and both holes residing in the same valley. As these three cases require one of the holes to reside in the lower spin-orbit split band in order for the biexciton to be bound, we do not consider these states. Of particular interest is the $A_1$ representation corresponding to both electrons and both holes residing in opposite valleys. Including the spin projections this corresponds to the following biexciton state, $B^{\uparrow K,\downarrow K'}_{\downarrow K,\uparrow K'}$ which is the semi-dark (due to momentum conservation) ground state singlet biexciton, and $B^{\uparrow K,\downarrow K'}_{\uparrow K,\downarrow K'}$  which is the excited bright state singlet biexciton. As the two states transform according to the same IrRep $A_1$, they can also be mixed by the electron-electron intervalley scattering process as in the trions case. The biexciton triplet states are given by $B^{\uparrow K,\downarrow K'}_{\uparrow K, \uparrow K'}$ and $B^{\uparrow K,\downarrow K'}_{\downarrow K, \downarrow K'}$ both being optically bright. The biexciton states transforming according to the $E_1$ IrRep are bright having both electrons in the same valley and both holes in opposite valleys, $[B^{\uparrow K, \downarrow K'}_{\uparrow K, \downarrow K}; B^{\uparrow K, \downarrow K'}_{\uparrow K', \downarrow K'}]$.
		
		\begin{table}
			\centering
			\captionsetup{labelfont=bf}
			\caption{\small{{\bf Group theory classification.}\\Summary of the group theory classification of excitonic complexes, $X$-excitons, $T$-trions, and $B$- Biexcitons, in Tungsten dichalcogenides according to the irreducible representations of the extended point group $C_{3v}''$}. The last column indicates the corresponding symbol used in the main text.}
			\label{summary}
			\begin{tabular}{||c|c|c|c|c|c||}
				\hline\hline
				& IrRep                   & States                                                                                                                                 & Bright     & Dark                  & \begin{tabular}[c]{@{}c@{}}Exciton or complex\\  (see Fig. 2)\end{tabular} \\ \hline
				\multirow{4}{*}{$X$} & \multirow{2}{*}{$E$}    & $[X^{\uparrow K}_{\downarrow K}; X^{\downarrow K'}_{\uparrow K'}]$                                                                     &            & \checkmark            & $X_d$                                                                           \\
				&                         & $[X^{\uparrow K}_{\uparrow K}; X^{\downarrow K'}_{\downarrow K'}]$                                                                     & \checkmark &                       & $X_b$                                                                           \\ \cline{2-6} 
				& \multirow{2}{*}{$E_3'$} & $[X^{\uparrow K}_{\uparrow K'}; X^{\downarrow K'}_{\downarrow K}]$                                                                     &            & \checkmark            & $X_d$                                                                           \\
				&                         & $[X^{\uparrow K}_{\downarrow K'}; X^{\downarrow K'}_{\uparrow K}]$                                                                     &            & \checkmark            &                                                                                 \\ \hline
				\multirow{6}{*}{$T$} & \multirow{4}{*}{$E_2'$} & $[T^{\uparrow K}_{\downarrow K, \uparrow K'}; T^{\downarrow K'}_{\downarrow K, \uparrow K'}]$         \rdelim\}{2}{8.5mm}[mix] &            & \checkmark            & $T_{sd}$                                                                        \\
				&                         & $[T^{\uparrow K}_{\uparrow K, \downarrow K'}; T^{\downarrow K'}_{\uparrow K, \downarrow K'}]$     \rdelim.{1}{8.5mm}[]         & \checkmark &                       & $T^*$                                                                           \\ \cline{3-6} 
				&                         & $[T^{\uparrow K}_{\uparrow K, \uparrow K'}; T^{\downarrow K'}_{\downarrow K, \downarrow K'}]$                                          & \checkmark &                       & $T$                                                                             \\
				&                         & $[T^{\uparrow K}_{\downarrow K, \downarrow K'}; T^{\downarrow K'}_{\uparrow K, \uparrow K'}]$                                          &            & \checkmark            & $-$                                                                                \\ \cline{2-6} 
				& $E$                     & $[T^{\uparrow K}_{\uparrow K, \downarrow K}; T^{\downarrow K'}_{\uparrow K', \downarrow K'}]$                                          & \checkmark &                       &        $T$                                                                          \\ \cline{2-6} 
				& $E_3'$                  & $[T^{\downarrow K'}_{\uparrow K, \downarrow K}; T^{\uparrow K}_{\uparrow K', \downarrow K}]$                                           &            & \checkmark            &            $-$                                                                     \\ \hline
				\multirow{5}{*}{$B$} & \multirow{4}{*}{$A_1$}  & $B^{\uparrow K,\downarrow K'}_{\downarrow K,\uparrow K'}$      \rdelim\}{2}{0mm}[mix]                                          &            & \checkmark            & $B_{sd}$                                                                        \\
				&                         & $B^{\uparrow K,\downarrow K'}_{\uparrow K,\downarrow K'}$         \rdelim.{1}{0mm}[]                                        & \checkmark &                       & $B^*$                                                                           \\ \cline{3-6} 
				&                         & $B^{\uparrow K,\downarrow K'}_{\uparrow K, \uparrow K'}$                                                                               & \checkmark &                       & $B$                                                                             \\
				&                         & $B^{\uparrow K,\downarrow K'}_{\downarrow K, \downarrow K'}$                                                                           & \checkmark & \multicolumn{1}{l|}{} & $B$                                                                             \\ \cline{2-6} 
				& $E_1'$                  & $[B^{\uparrow K, \downarrow K'}_{\uparrow K, \downarrow K}; B^{\uparrow K, \downarrow K'}_{\uparrow K', \downarrow K'}]$               & \checkmark &                       & $B$                                                                             \\ \cline{2-6} 
				\hline\hline
			\end{tabular}
		\end{table}
		
		\clearpage
		\newpage

		\section{Model calculations of the intervalley scattering matrix element}
		\label{sec:mat}
		
		\subsection{Ab initio density functional theory}

		In the DFT calculations the wave functions were obtained in the local density approximation,
		using a plane-wave basis of $600~{\rm eV}$ cutoff energy and a k-point
		grid of $12 \times 12$ in the 2D Brillouin zone. We used the VASP\cite{vasps} code for these calculations, which employs periodic boundary conditions in three dimensions even for 2D materials; for this reason we used a large inter-layer distance of $20~{\rm \AA}$ to mimic the limit of an isolated monolayer. The form factor was calculated by post-processing the DFT wave functions, simply taking the matrix element of the bare Coulomb interaction between the initial and final states of the scattering process. In the calculation of this matrix element we neglected spin-orbit coupling.
		
		The form factor was calculated in reciprocal space by Fourier 
		transforming Eq.~(2) in the main text, leading to a summation on the 
		grid of reciprocal lattice vectors. This technique is sensitive to the 
		plane-wave cutoff energy. We have therefore  tested the sensitivity of 
		the form factor to the cutoff energy by calculating it for WS$_2$ with 
		an extremely reduced cutoff of $100~{\rm eV}$ and an increased cutoff of $900~{\rm eV}$. 
		We found that reducing the cutoff reduces the form factor by $10~\%$, 
		while increasing the cutoff increases the form factor by $3~\%$. 
		
		Convergence of the calculation was also tested for the inter-layer 
		separation. We found that decreasing the separation to $15~{\rm \AA}$ only 
		changes the form factors by less than $1~\%$.
		
		In Fig.~S1 we show the DFT calculated band structure for ${\rm WS_2}$ and ${\rm WSe_2}$, showing the band edges at the $K$ point and the spin-orbit splitting.
		In Tables~S4 and S5 we list the DFT obtained orbital decomposition of the electron states at the $K/K'$ points in the conduction and valence bands demonstrating the dominance of the transition metal $d$ orbitals.
		
		\begin{figure}[!h]
			\centering
			\includegraphics[width=1\textwidth]{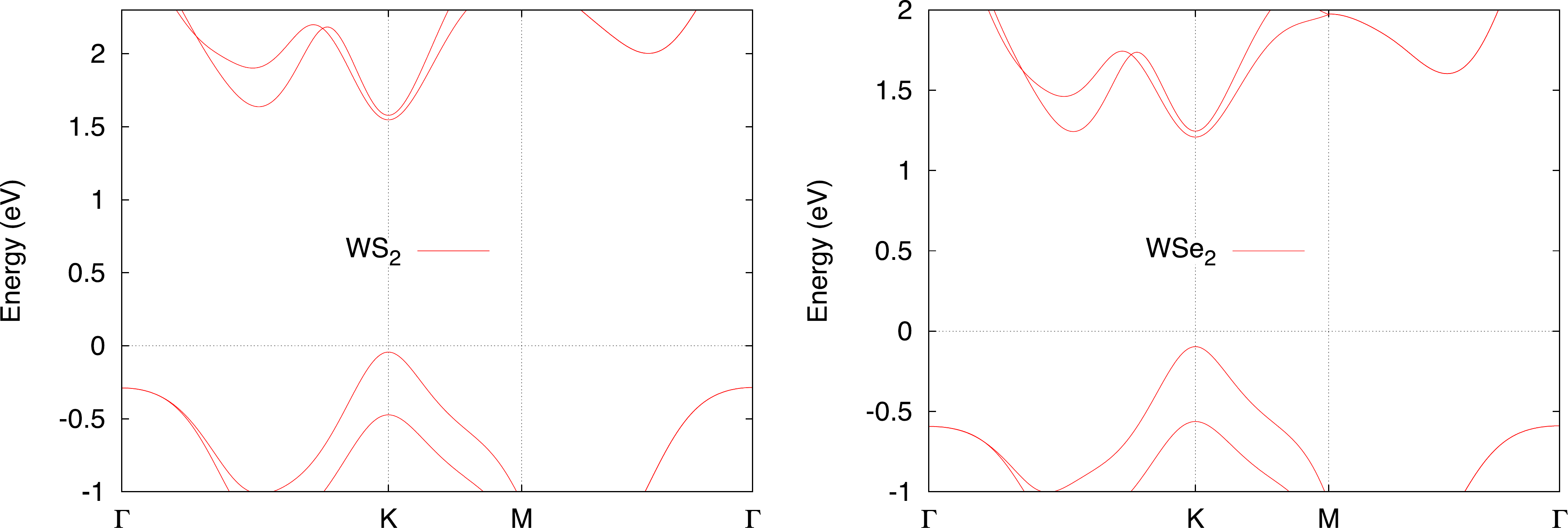}
			\captionsetup{labelfont=bf}
			\captionsetup{labelsep=period-newline}
			\caption{ 
				{\bf DFT calculated band structure of ${\rm WX_2}$}.\\
			}
		\vspace{0.5cm}
			\label{fig:conv}
		\end{figure}
		\begin{table}[!h]
			\centering
			\captionsetup{labelfont=bf}
			\captionsetup{labelsep=period-newline}
			\caption{\small{{\bf DFT calculated orbital decomposition at the $K/K'$ point in ${\rm WS_2}$.}\\
			}}
			\label{dft-decomp}
			\begin{tabular}{cccccccc}
				\hline\hline
				band & ${\rm W}-5d_{z^2}$ & ${\rm W}-5d_{x^2-y^2}$ & ${\rm W}-5d_{xy}$ & ${\rm W}-6s$ & ${\rm S}-p_x$ & ${\rm S}-p_y$  & \\
				\hline
				$c$ & $86.9\%$ & $0$ & $0$ & $7.8\%$ & $2.6\%$ & $2.6\%$  \\ 
				$v$ & $0$ &  $39.5\%$ & $39.5\%$ & $0$  & $10.2\%$  & $10.2\%$ \\ 
				\hline \hline
			\end{tabular} 
		\end{table}

		\begin{table}[!h]
			\centering
			\captionsetup{labelfont=bf}
			\captionsetup{labelsep=period-newline}
			\caption{\small{{\bf DFT calculated orbital decomposition at the $K/K'$ point in ${\rm WSe_2}$.}
			}}
			\label{dft-decomp}
			\begin{tabular}{cccccccc}
				\hline\hline
				band & ${\rm W}-5d_{z^2}$ & ${\rm W}-5d_{x^2-y^2}$ & ${\rm W}-5d_{xy}$ & ${\rm W}-6s$ & ${\rm Se}-p_x$ & ${\rm Se}-p_y$  & \\
				\hline
				$c$ & $85.9\%$ & $0$ & $0$ & $8.1\%$ & $2.2\%$ & $2.2\%$  \\ 
				$v$ & $0$ &  $40.1\%$ & $40.1\%$ & $0$  & $9.2\%$  & $9.2\%$ \\ 
				\hline \hline
			\end{tabular} 
		\vspace{1cm}
		\end{table}
		
		\subsection{Tight-binding model}
		In the tight binding model, the Bloch wave function of the conduction band electrons at the $K$ point, using only the transition metal $d$-orbital  is given by
		\begin{equation}
		\Psi(\vec{r})=\frac{C}{\sqrt{N}}\sum_{i}e^{i\vec{K}\cdot \vec{R}_i}\phi(\vec{r}-\vec{R}_i),
		\end{equation}
		where $N$ is the number of unit cells, $\vec{R}_i$ is the lattice vector coinciding with the transition metal atoms positions, and $C$ is the weight of the $5d_{z^2}$ orbital $\phi$ centred on $\vec{R}_i$. The value of $C$ is obtained from the orbital decomposition given in Tables~S4, S5 for ${\rm WS_2}$ and ${\rm WSe_2}$, respectively.
		The 3D coulomb matrix element is given by
		\begin{equation}
		M=e^2\int \frac{d^3\vec{r}_1d^3\vec{r}_2}{|\vec{r}_2-\vec{r}_1|}\Psi^*(\vec{r}_1)\Psi^*(\vec{r}_2)\Psi(\vec{r}_1)\Psi(\vec{r}_2).
		\end{equation}
		Plugging in the Bloch wave function and using the two-centre approximation for the electron-electron Coulomb interaction we get 
		\begin{equation}
		M=e^2|C|^4\sum_{\vec{R}} e^{i\vec{K}\cdot\vec{R}}\int d^3\vec{r}_1d^3\vec{r}_2\frac{|\phi(\vec{r}_1)|^2|\phi(\vec{r}_2)|^2}{|\vec{r}_2-\vec{r}_1+\vec{R}|},
		\end{equation}
		where the summation is over the lattice sites $\vec{R}=l\vec{a}_1+n\vec{a}_2$, where $\vec{a}_1=a_0(1,0)$, and $\vec{a}_2=\frac{a_0}{2}(1,\sqrt{3})$ are the lattice primitive vectors, $a_0$ is the lattice constant, and $l, n$ are integers. Finally, the matrix element is related to the dimensionless parameter $\chi$ through the intervalley interaction Hamiltonian giving,
		\begin{equation}
		\chi=\frac{m_c}{m}\frac{A}{a_B}|C|^4\sum_{\vec{R}}e^{i\vec{K}\cdot\vec{R}}\int d^3\vec{r}_1d^3\vec{r}_2\frac{|\phi(\vec{r}_1)|^2|\phi(\vec{r}_2)|^2}{|\vec{r}_2-\vec{r}_1+\vec{R}|},
		\label{eq:mat}
		\end{equation}
		where $m_c$ is the $c$-band electron mass, $m$ is the free electron mass, $A$ is the unit cell area, and $a_B$ is the Bohr radius.
		
		For the atomic orbital entering into the Coulomb matrix element we use the Roothaan-Hartree-Fock (RHF) atomic orbitals\cite{rhfs,wang_rhfs} which consist of a linear combination of Slater-type orbitals,
		\begin{align}
		\phi_{nlm}(\vec{r})=Y^l_m(\theta,\phi)\sum_j C_{j} S_{j}(r)=Y^l_m(\theta,\phi)R_{nl}(r),
		\end{align}
		where $n, l, $ and $m$ are the principle, azimuthal and magnetic quantum numbers, and $Y^l_m(\theta,\phi)$ are the spherical harmonics.
		The Slater-type radial orbital $S(r)$ has the general form
		\begin{equation}
		S(r)=N_s r^{n-1} e^{-Zr},
		\end{equation}
		here $N_s=\frac{(2Z)^{n+1/2}}{\sqrt{(2n)!}}$ is a normalization constant, and $Z$ is the orbital exponent. Using the tables in Ref. [\citenum{rhfs}] we construct the Tungsten $5d_{z^2}$ orbital, with the radial part given by (in atomic units)
		\begin{equation}
		\begin{split}
		&R_{5d}(r)=-1070.29 e^{-29.4731 r} r^2 - 1297.24 e^{-18.363 r} r^2
		\\
		& + 1192.26 e^{-12.073 r} r^3 + 239.385 e^{-7.9781 r} r^3
		\\
		& -56.2785 e^{-5.19312 r} r^4 - 7.74766 e^{-3.14551 r} r^4
		\\
		& - 0.18956 e^{-1.79159 r} r^4,
		\end{split}
		\end{equation}
		and the angular part is $Y^2_0(\theta,\phi)=\frac{\sqrt{5}}{4\pi}(3\cos^2\theta-1)$.
		
		We separate the calculation of the matrix element into two parts, first taking $\vec{R}=0$ giving the on-site contribution, and then allowing for $\vec{R}\ne 0$.
		For the on-site contribution with $\vec{R}=0$, we expand the Coulomb potential in spherical harmonics
		\begin{align}
		\frac{1}{|\vec{r}_2-\vec{r}_1|}=\sum_{l=0}^{\infty}\frac{r^l_<}{r_>^{l+1}}
		\sum_{m=-l}^{m=l}\frac{4\pi}{2l+1}{Y^l_m}^*(\theta',\phi')Y^l_m(\theta,\phi),
		\label{eq:sph}
		\end{align}
		which allows to separate the radial and angular integrations.
		The angular integration consists of products of three spherical harmonics which can be written in terms of Wigner 3j-symbols,
		\begin{align}
		&\int Y^{l_1}_{m_1}(\theta,\phi)Y^{l_2}_{m_2}(\theta,\phi)
		Y^{l_3}_{m_3}(\theta,\phi)\sin \theta d\theta d\phi
		\\
		&=\sqrt{\frac{(2l_1+1)(2l_2+1)(2l_3+1)}{4\pi}}
		\left(\begin{matrix}
		l_1 & l_2 & l_3 \\
		0 & 0 & 0 
		\end{matrix}\right)
		\left(\begin{matrix}
		l_1 & l_2 & l_3 \\
		m_1 & m_2 & m_3
		\end{matrix}\right). \nonumber
		\end{align}
		The Wigner 3j-symbols impose selection rules on the possible values of the different angular momentum quantum numbers, thus reducing the number of terms in the sum and the number of integrations needed. In particular we must have, $m_1+m_2+m_3=0, |m_i|<l_i$, and $|l_1-l_2|\le l_3\le l_1+l_2$.
		
		For the case of non-zero $\vec{R}$, since the wave functions have a typical spread smaller than the lattice constant, we use the following expansion\cite{bipolars,twocenters} valid for $|\vec{r}_1+\vec{r}_2|<R$,
		\begin{figure}
			\centering
			\includegraphics[width=0.7\textwidth]{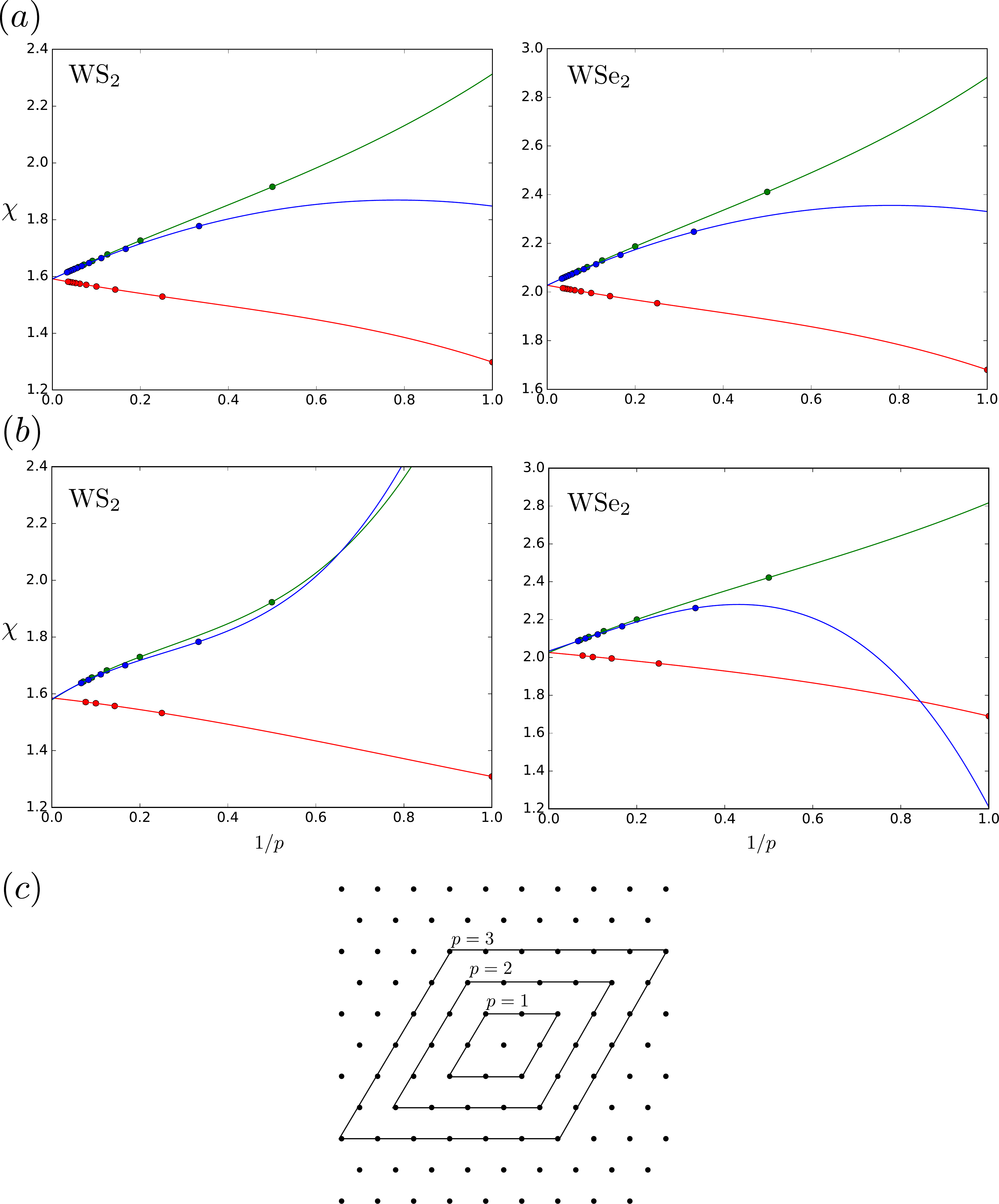}
			\captionsetup{labelfont=bf}
			\captionsetup{labelsep=period-newline}
			\caption{ 
				{\bf Convergence of the intervalley scattering matrix element calculation}.\\
				(a) Analytical calculation of the matrix element as a function of the inverse number of lattice points in the summation. (b) Monte Carlo calculation results.
				We fit the points to third order polynomials and extract the value for $1/p\rightarrow0$ corresponding to summation over an infinite lattice. The data points are separated into three sequences with a period of 3, all converging to the same point. This behaviour of the sum is attributed to the phase factor in the summation involving the $\vec{K}$ vector, and to the rhombic unit cell used in the summation.
				(c) Sketch of the rhombic unit cell used for the summation over the triangular lattice points for increasing values of $p$.}
			\label{fig:conv}
		\end{figure}
		\begin{align}
		&\frac{1}{|\vec{r}_2-\vec{r}_1+\vec{R}|}=
		\sum_{l_a,l_b=0}^{\infty} R^{-(l_a+l_b+1)}r_1^{l_a}r_2^{l_b}V_{l_a,l_b};
		\\
		&V_{l_a,l_b}=(4\pi)^{3/2}(-1)^{l_b}\left(\begin{matrix}
		2(l_a+l_b) \\ 2l_a
		\end{matrix}\right)^{1/2} \nonumber
		\\
		&\times[(2l_a+1)(2l_b+1)(2(l_a+l_b)+1)]^{-1/2} \nonumber
		\\
		&\times\sum_{M=-(l_a+l_b)}^{l_a+l_b}(-1)^M Y^{L}_{-M}(\hat{R})[Y^{l_a}(\hat{r}_1)\otimes Y^{l_b}(\hat{r}_2)]_M^{l_a+l_b}; \nonumber
		\\
		&[Y^{l_a}(\hat{r}_1)\otimes Y^{l_b}(\hat{r}_2)]^{l_a+l_b}_{M}=\sum_{m_a=-l_a}^{l_a}\sum_{m_b=-l_b}^{l_b}Y^{l_a}_{m_a}(\hat{r}_1)Y^{l_b}_{m_b}(\hat{r}_2) \nonumber
		\\
		&\times \langle l_am_a;l_bm_b|(l_a+l_b)M\rangle. \nonumber
		\end{align}
		
		In Fig.~S2 we show the convergence of the summation using both the detailed analytical method and a Monte Carlo calculation of the integral in Eq.~({\ref{eq:mat}), showing that both methods converge to the same value for the dimensionless matrix element $\chi$.
		
	}
	\clearpage
	\bibliographystyle{naturemag}
	


\end{document}